\documentclass{llncs}

\newcommand{\struc}{\mathcal{I}}
\usepackage{color}
\definecolor{lightgray}{rgb}{.9,.9,.9}
\definecolor{darkgray}{rgb}{.4,.4,.4}
\definecolor{purple}{rgb}{0.65, 0.12, 0.82}

\usepackage{verbatimbox}

\usepackage{wrapfig}

\usepackage{amsmath}

\usepackage{listings}
\author{Joost Vennekens}
\institute{Dept.~Computer Science, KU Leuven\\
Technology Campus De Nayer, Sint-Katelijne-Waver, Belgium
}
\usepackage[hyphens]{url}
\title{Lowering the learning curve for declarative programming: a Python API for the IDP system}

\begin{document}

\maketitle

\begin{abstract}
Programmers may be hesitant to use declarative systems, because of the associated learning curve. In this paper, we present an API that integrates the IDP Knowledge Base system into the Python programming language. IDP is a state-of-the-art logical system, which uses SAT, SMT, Logic Programming and Answer Set Programming technology. Python is currently one of the most widely used (teaching) languages for programming. The first goal of our API is to allow a Python programmer to use the declarative power of IDP, without needing to learn any new syntax or semantics. The second goal is allow IDP to be added to/removed from an existing code base with minimal changes.
\end{abstract}

\section{Introduction}

While declarative systems may not be inherently more difficult to learn
than, e.g., an object-oriented programming language, they are still
often perceived as such, because many programmers are unfamiliar
with their syntax, semantics or paradigm. This leads in turn to a chicken-and-egg problem, where programmers do not learn this technology, because companies
are reluctant to adopt it, because there are not enough programmers
who know the technology readily available.

In this paper, we propose to tackle both problems by means of an API that allows a declarative knowledge base (KB) to be used from within a
well-known imperative host language.  Our first goal is
to integrate KB functionality into the host language
as seamlessly as possible. In this way, it should be possible to use the
knowledge base to prototype a single component of a large system, without affecting the rest of the code base. The second goal is to have a very low learning curve for the API. We achieve this by using as much as possible the syntax of the host language, and by requiring no more background in declarative languages than an introductory course on classical logic. We therefore expect our API to be immediately usable by, e.g., bachelor students in a typical CS curriculum.

In more detail, we postulate these guidelines for our development of the API:
\begin{itemize}
\item The interaction between the KB and host language should be done through standard objects of the host language.
\item The need to learn KB-specific terminology should be kept to a minimum.
\item It should be as easy as possible to replace the KB by a piece of host language code (or, vice versa, to replace a piece of host language code by a KB).
\end{itemize}

A typical use for our API will be to off-load specific computational problems (e.g., detect connected components in a graph, find a permissible  allocation of resources to jobs) to the KB, thereby avoiding the need to implement a specific algorithm and thus arriving at a working prototype more quickly.  In such an early prototype, the modular, declarative nature of the KB will be particularly useful, because of its ability to easily cope with additional changes to the specification. Once the program has reached a certain level of maturity, it can of course be profiled to see whether all of the KB components meet the performance requirements. Whenever this is not the case, the KB can be replaced by a dedicated algorithm with minimal impact on the rest of the code.

As a host language, we use Python (in particular, version 2.7).  Given our stated goals, this is the most obvious choice: ``[a]t the time of writing (July 2014), Python is currently the most popular language for teaching introductory computer science courses at top-ranked U.S. departments.''\footnote{\url{http://cacm.acm.org/blogs/blog-cacm/176450-python-is-now-the-most-popular-introductory-teaching-language-at-top-us-universities/fulltext}} We assume familiarity with the basics of Python throughout this paper.  The KB system that we use will be discussed in Section \ref{sec:kbs}. Section \ref{sec:int} then discusses our interface between host language and KB, which we validate by means of 
some examples in Section \ref{sec:ex}. In Section \ref{sec:impl}, we give some brief notes on the implementation of our API.  Finally, Section \ref{sec:rel} discusses some related work, in particular, other approaches that integrate a declarative knowledge base into an imperative language.

\section{KB system}\label{sec:kbs}

As an underlying KB system, we will use IDP ({\em Imperative Declarative Programming})\footnote{\url{https://dtai.cs.kuleuven.be/software/idp}}  \cite{bruynooghe14}, which combines techniques from SAT solving, Logic Programming and Answer Set Programming (ASP).  It has performed well in previous ASP competitions, e.g., narrowly finishing second after Clasp in the System Track of 2011\footnote{\url{https://www.mat.unical.it/aspcomp2011/}}.  IDP has a number of properties that fit well with our goals of achieving both a tight integration with the host language and a low learning curve.

{\em Input language.} IDP uses a language that is a conservative extension of classical first-order logic (FO).  Because most students of computer science are familiar with FO, this means that the learning curve for a large part of IDP's input language consists only of learning a particular ASCII syntax for FO. One of the ways in which IDP extends FO is by adding {\em inductive definitions} \cite{denecker08}. Because such definitions cannot, in general, be expressed in FO, this is a real extension of the language. Moreover, since inductive datatypes (lists, trees, \ldots) are very common in computer programs, this feature will prove useful in our API.

{\em Inference.} In addition to its input language, a second useful property of IDP is that it aims to support a variety of different inference tasks.  Particularly useful in the context of  this article is the task of {\em finite model expansion}. As pointed out in \cite{MitchellT05}, modal expansion for FO captures the complexity class NP, thereby covering the kind of tasks that we would like to off-load to a declarative KB. Moreover, \cite{Tasharrofi} have further demonstrated that model expansion is a key task when using declarative methods to build modular software systems.

\subsection{FO: syntax and semantics}

\renewcommand{\struc}{S}

We briefly recall the standard syntax and semantics of FO. A {\em vocabulary} $\Sigma$ consists of a set of function symbols, each with an associated arity $n$, and a set of predicate symbols, also each with an arity $n$. A function with arity 0 is called a {\em constant}. A {\em term} is either a constant, a variable or an expression $f(t_1,\ldots,t_n)$ where $f$ is an $n$-ary function symbol and the $t_i$ are terms. An {\em atom} is an expression $P(t_1,\ldots,t_n)$, with $P$ an $n$-ary predicate and the $t_i$ again terms.  A {\em formula} is either an atom or an expression $\psi \lor \phi$, $\psi \land \phi$, $\lnot \phi$, $\forall x: \phi$, or $\exists x: \phi$, where $\psi,\phi$ are formulas and $x$ is a variable. As usual, $\phi \Rightarrow \psi$ abbreviates $\lnot \phi \lor \psi$ and $\phi \Leftrightarrow \psi$ stands for $(\phi \Rightarrow\psi)\land (\psi \Rightarrow\phi)$. A {\em sentence} is a formula without free variables and a {\em theory} is a finite set of sentences.

The standard semantics of FO is defined in terms of {\em structures} for a vocabulary $\Sigma$. Each structure $\struc$ consists of a domain $D$ and a mapping of:
\begin{itemize}
\item Each $n$-ary predicate symbol $P$ in $\Sigma$ to an $n$-ary relation $R \subseteq D^n$
\item Each $n$-ary function symbol $f$ in $\Sigma$ to an $n$-ary function $f: D^n \rightarrow D$
\end{itemize}
The {\em satisfaction} relation $\models$ is defined between structures for a vocabulary and theories of this  vocabulary  (or a subvocabulary thereof) by the usual induction. When $\struc \models T$, we also say that the structure $\struc$ is a {\em model} of $T$.

\subsection{FO in the IDP system}\label{sec:idp}

\begin{wrapfigure}[8]{r}{5cm}
\centering
\begin{tabular}{c|c|c}
FO & IDP & Python \\
\hline
$\land $& \verb|&| & \verb|and|\\
$\lor $& \verb@|@ & \verb|or|\\
$\lnot $& \verb|~| & \verb|not|\\
$\Rightarrow $& \verb|=>| & not present  \\
$= $& \verb|=| & \verb|==|\\
$\neq $& \verb|~=| & \verb|!=| \\
$\forall $& \verb|!| & \verb|all|\\
$\exists $& \verb|?| & \verb|any| \\
\end{tabular}
\caption{ASCII syntax of IDP.\label{fig:syntax}}
\end{wrapfigure}

IDP uses the standard concepts of vocabularies, theories and structures. Each of these has a specific syntactic representation. For example, a map coloring problem can be described in the following vocabulary $V$.

\begin{verbatim}
vocabulary V {
    type Color
    type Area
    Border(Area,Area)
    Coloring(Area): Color
}
\end{verbatim}


As can be seen here, IDP in fact uses a typed variant of first-order logic.  The first two statements define two types (which can be seen as unary predicates), whereas the last two statements define, respectively, a predicate and a function.

The following  theory $T$ in vocabulary $V$ consists of a single sentence, expressing that neighboring areas must have a different color. Fig.~\ref{fig:syntax} shows the ASCII symbols that are used in IDP to represent the logical connectives.
\begin{verbatim}
theory T : V {
    !a b: Border(a,b) => Coloring(a) ~= Coloring(b).
}
\end{verbatim}

IDP is usually able to automatically derive the types of variables, based on the type declarations in the definition of the vocabulary. This information is important because most of its inference tasks require IDP to first {\em ground} (part of) the theory.  Instead of depending on IDP's automatic type derivation, it is also possible to explicitly declare the type of a variable, e.g.:
\begin{verbatim}
!a [Area] b [Area]: Border(a,b) => Coloring(a) ~= Coloring(b).
\end{verbatim}

A structure $S$ for a vocabulary $V$ is represented by an enumeration (1) of the values that belong to each type, (2) of the tuples that belong to each predicate, and (3) of the mapping of tuples to values that is made by each function.

\begin{verbatim}
structure S : V {
  Area  = { Belgium; Holland; Germany; }
  Color = { Blue; Red; Green; }
  Border = {(Belgium,Holland); (Belgium,Germany); (Holland,Germany)}
}
\end{verbatim}
This structure $S$ for the vocabulary $V$ interprets only part of the vocabulary $V$. In particular, the function $Coloring$ is not interpreted. One of the inference tasks supported by the IDP system is that of {\em model expansion}: given a structure $S'$ for a subvocabulary $\Sigma'$ of the vocabulary $\Sigma$ of a theory $T$, compute a structure $S$ for the remaining symbols $\Sigma \setminus \Sigma'$ such that $(S' \cup S) \models T$.

IDP exposes its functionality by means of an API in the Lua scripting language. The command \verb|printmodels(modelexpand(T,S))| performs the model expansion task for the above structure $S$ and theory $T$, resulting in the output:
\begin{verbatim}
Coloring = {"Belgium"->"Red";"Germany"->"Blue";"Holland"->"Green"}
\end{verbatim}
By default, a single model expansion is computed, but it is also possible to compute several or all of them. A special case of model expansion occurs when the initial structure $S$ already interprets the entire vocabulary $V$ of the theory $T$. In this case,  it reduces to checking whether $S \models T$.

\section{Interfacing with the KB System} \label{sec:int}

This section presents our API for using the IDP KB system from within Python.

\subsection{Vocabularies and structures} 

All interaction with the IDP system is done through objects of the \verb|IDP| class.  Each such object represents a knowledge base consisting of a triple $(V,S,T)$ of a vocabulary $V$, a structure $S$ and a theory $T$. It can be created as follows:

\begin{lstlisting}
kb = IDP()
\end{lstlisting}

The following methods add symbols to the vocabulary of the KB:
\begin{lstlisting}
kb.Type(name [, interpretation])
kb.Constant(typed_name [, interpretation]
kb.Function(typed_name [, interpretation])
kb.Predicate(typed_name [, interpretation])
\end{lstlisting}

As in IDP itself, the \verb|typed_name| of a predicate is a string of the form 
\verb|Foo(Type1, ..., Type2)| and that of a
function is  \verb|Foo(Type1, ..., Type2):| \verb|Return_type|.  Because constants are identical to $0$-ary functions,
their \verb|typed_name| has the form \verb|Foo : Type|.  Once
a symbol \verb|Foo| has been added to the vocabulary of a knowledge base
\verb|kb|, it can thereafter be referred to as \verb|kb.Foo|.

In addition to declaring a function/predicate  symbol $\sigma$ (i.e., adding it to the vocabulary of the KB),
it is also possible to immediately extend the structure $S$ with
a particular interpretation $\sigma^S$ for $\sigma$. This is done by adding this interpretation as a second
argument.  The interpretation of a type must be a
set (or list) of values; that of a constant must be a
single value; that of a function with arity $\geq 1$ must be mapping (e.g., a dictionary); and that of a predicate must be a set/list of tuples of the correct arity (for predicates with arity $1$, a set of simple values is also allowed). Obviously, the typing of the symbols must be respected.

Instead of initialising the interpretation of a symbol upon construction, it is also possible to first declare a symbol and then later use the assignment operator to provide an interpretation for it. 
We illustrate using (part of) the graph coloring example of Sec.~\ref{sec:idp}.

\begin{lstlisting}
color = IDP()
color.Type("Color", ["Blue", "Red", "Green"])
color.Type("Area",  ["Belgium","Holland", "Germany"]) 
color.Predicate("Border(Area,Area)")
color.Border=[("Belgium","Holland"),("Belgium","Germany"), 
              ("Holland","Germany")]
color.Function("Coloring(Area): Color")
\end{lstlisting}

The ``logical'' objects that are thus created implement a number of common Python interfaces, allowing them to act as Python programmers would expect.
A {\em relation} is, in mathematical terms, a set of tuples.  Its natural
counterpart is a Python {\em MutableSet} object\footnote{\url{https://docs.python.org/2/library/collections.html#collections.Set}}
  (i.e., a set which allows adding/removing of elements).  The following interactive session demonstrates some standard usages.
\begin{lstlisting}
>>> "Belgium" in color.Area
True
>>> color.Area.add("Austria")
>>> for x in color.Area:
...     print x,
... 
Holland Austria Germany Belgium
\end{lstlisting}
 In addition to the standard {\em MutableSet} functionality, relations are also  {\em callable}\footnote{\url{https://docs.python.org/2/library/functions.html#callable}}, so that we may also use the standard FO notation for checking membership.
\begin{lstlisting}
>>> color.Area("Belgium")
True
\end{lstlisting}

A  {\em function} is a {\em
  Mapping}\footnote{\url{https://docs.python.org/2/library/collections.html#collections.Mapping}},
that is, an object that maps each tuple of values in its domain to a
value in its range.  Some standard usages are:
\begin{lstlisting}
>>> color.Coloring.keys()
['Austria', 'Germany', 'Holland', 'Belgium']
>>> color.Coloring["Belgium"]
'Blue'
\end{lstlisting}
As with predicates, function objects are 
 {\em callable} to allow for the more FO-like:
\begin{lstlisting}
>>> color.Coloring("Belgium")
'Blue'
\end{lstlisting}

\subsection{Formulas and definitions}

In keeping with our goal of achieving a low learning curve, formulas are written in Python syntax.  An overview is shown in Fig.~\ref{fig:syntax}.  The Python language has the standard boolean operators \verb|and|, \verb|or| and \verb|not|.  In addition, it also has the functions \verb|all| and \verb|any|, which may be applied to lists of boolean values to return the conjunction/disjunction of these  values. The latter two functions, together with Python's list comprehension syntax, can be used as universal/existential quantification.  The list comprehension syntax also has an optional \verb|if| part, which may be used to represent the common pattern of a universally quantified implication:

\begin{center}
\begin{tabular}{ccc}
$\forall x[Type]: \phi$ &\hspace{2cm} & \verb|all(|$\phi$\verb| for x in Type)| \\
$\exists x[Type]: \phi$ && \verb|any(|$\phi$\verb| for x in Type)| \\
$\forall x[Type]: \phi \Rightarrow \psi$ && \verb|all(|$\psi$\verb| for x in Type if |$\phi$\verb|)|
\end{tabular}
\end{center}

In the graph coloring problem, we need to express the following property:
\begin{equation}\label{eq:graph}
\forall a\ b: Border(a,b) \Rightarrow Coloring(a) \neq Coloring(b)
\end{equation}
Section \ref{sec:idp} already presented the IDP syntax for this. In Python, we can write the same property as:
\begin{lstlisting}
all(color.Coloring(a) != color.Coloring(b) 
                      for (a,b) in color.Border)
\end{lstlisting}
Note that this is just a normal Python expression, which we can, e.g., just type into the interactive terminal. This expression evaluates to \verb|True| precisely when property \eqref{eq:graph} is satisfied.  We can make the KB aware of such a constraint by means of its \verb|Constraint| method, which takes a string as its argument.
So, the following code adds the above constraint to our graph coloring KB:

\begin{lstlisting}
color.Constraint("all(Coloring(a) != Coloring(b) 
                                  for (a,b) in Border")
\end{lstlisting}

The string argument is completely identical to the Python expression we saw above, with one exception: the predicates/functions are simply called \verb|Coloring|, \verb|Area| and \verb|Border|, instead of \verb|color.Coloring|, \verb|color.Area| and \verb|color.Border|. This is because, just as a theory can only contain symbols that appear in its vocabulary, the constraints that are added to a KB always refer to the symbols of that KB.

An alternative formula, which is equivalent to the one above, quantifies over $Area \times Area$ and uses an \verb|if|-expression to check for membership in $Border$:
\begin{lstlisting}
color.Constraint("all(Coloring(a) != Coloring(b) 
      for a in Area for b in Area if (a,b) in Border)")
\end{lstlisting}




\subsection{Functional interfacing}

In keeping with our goal of making the API easy to use, the programmer does not need to explicitly invoke the IDP system.  This avoids the need to learn new functions or new terminology, and reduces the possibility of bugs.  Instead, invokation of the IDP system happens ``automagically''  in the following circumstances:
\begin{itemize}
\item Symbols that have been declared, but for which no interpretation has been provided, are automatically assigned a valid interpretation (in accordance with FO semantics) when their content is inspected. In other words, IDP is used as an {\em oracle} to lazily fill in the interpretation of any declared symbols for which the user does not provide one herself. This is done in a way that the interpretations of all symbols together constitutes an FO model of the constraints, i.e., a model expansion task is performed. If the constraints admit multiple models, one is chosen arbitrarily.
\item The KB object has an attribute \verb|satisfiable|, which is automatically set to \verb|True|/\verb|False| (depending on whether the KB is satisfiable) when the user converts it to a boolean (either explicitly with \verb|bool(.)| or by use in an \verb|if|-statement).
\end{itemize}

In the previous section, we declared a function \verb|color.Coloring|  without adding an interpretation for this function. Therefore, the IDP system will be invoked to compute a coloring of our graph if we execute the following code:
\begin{lstlisting}
for x in color.Area:
    print "Area %s has color %s" % (x, color.Coloring[x])    
\end{lstlisting}
Note that if we {\em had} added an interpretation for the function \verb|Coloring| before executing this code, then this \verb|for|-loop would still continue work as expected.
This is an important property, because it allows us to change whether a relation/function is computed by the KB base or by native Python code, without having to adjust the code that makes use of it.  In the latter case, the call \verb|color.Coloring[x]| will just retrieve the pre-computed coloring stored within the \verb|color| object, without invoking IDP.

Similarly, if a coloring is not provided, then the following code will test whether a given graph {\em can be} colored, whereas if it is provided, the same code checks whether it is indeed valid.
\begin{lstlisting}
if color.satisfiable:
   print "The graph can be colored"
\end{lstlisting}

\subsection{Inductive definitions}

An important feature of the IDP system is its ability to handle {\em inductive definitions}. It uses a rule-based syntax for representing such definitions, in which, e.g., the transitive closure $TC$ of a graph $G$ can be defined as follows

\[\left\{
\begin{aligned}
\forall x,y: TC(x,y)&\leftarrow G(x,y).\\
\forall x,y: TC(x,y)&\leftarrow \exists z:\ G(x,z) \land TC(z,y).\\
\end{aligned}
\right\}\]

Note that the arrow symbol here is not material implication, but a special symbol that denotes a ``case'' in an inductive definition.
Such an inductive definition is interpreted under the well-founded semantics \cite{vrs91}, which in the case of a positive definition (such as the one above) boils down to a least-fixpoint construction.  Each rule of such a definition represents a single case in which the defined predicate holds. In our Python API, we use a lambda-expression to represent such a case. 

\begin{lstlisting}
kb.Define([("TC(Node,Node)", "lambda x,y: G(x,y)"),
           ("TC(Node,Node)", "lambda x,y: 
                 any(G(x,z) and TC(z,y) for z in Node)")
\end{lstlisting}
This  both declares the predicate \verb|TC| and defines it in terms of the ``parameter'' \verb|G|.  For definitions consisting of a single rule,  a simpler syntax is also allowed:
\begin{lstlisting}
kb.Define("TC(Node,Node)", 
                "lambda x,y: G(x,y) or
                 any(G(x,z) and TC(z,y) for z in Node)")
\end{lstlisting}

Similar to how an argument of \verb|kb.Constraint(.)| can also be used as a simple boolean Python expression, the above lambda-expression can be used to compute the transitive closure of $G$ by an explicit least-fixpoint computation: 

\begin{lstlisting}
def lfp(f, x=[]):
    y = f(x)
    return y if y == s else f(x)

node_pairs = [(x,y) for x in kb.Node for y in kb.Node]
TC = lfp(lambda T: filter(lambda x,y: kb.G(x,y) or 
   any(kb.G(x,z) and T(z,y) for z in Node), node_pairs))
\end{lstlisting}

An advantage of using IDP is that the definition of $TC$ can then not only be used to compute the transitive closure of a given graph, but also to, e.g., compute a graph that  would have a given relation as its transitive closure. In addition, IDP not only supports positive inductive definitions, but also non-monotone inductive definitions (such as the standard definition of the relation ``$\models$'' in FO), for which a simple least-fixpoint construction does not work.  Non-recursive definitions (which are equivalent to a standard FO equivalence) are also allowed in IDP. In the latter case, we can of course choose whether to use the \verb|Define| or \verb|Predicate| method of our API. 

IDP can be configured to use XSB Prolog\footnote{\url{http://xsb.sourceforge.net/}} to speed up certain computation with definitions. We always use this option in the experiments below.

\section{Experiments} \label{sec:ex}

This section presents two examples of our API,  with a particular focus on demonstrating that the integration into the surrounding Python code can be done in a  natural way.

\subsection{Sudoku}

The first example is a Sudoku solver. 
A Sudoku grid consists of $9\times 9 = 81$ cells. 
\begin{lstlisting}
sud = KB()
sud.Type("Cell", range(81))
\end{lstlisting}

The grid is divided into in rows, columns and nine small $3 \times 3$ squares.
\begin{lstlisting}
sud.Predicate("SameRow(Cell, Cell)", 
    [ (i, j) for i in sud.Cell for j in sud.Cell 
      if i / 9 == j / 9 ])
sud.Predicate("SameCol(Cell, Cell)", 
    [ (i, j) for i in sud.Cell for j in sud.Cell  
       if i % 9 == j % 9 ])
sud.Predicate("SameSmallSq(Cell, Cell)", 
    [ (i, j) for i in sud.Cell  for j in sud.Cell 
       if (i%9)/3 == (j%9)/3 and (i/9)/3 == (j/9)/3 ])
\end{lstlisting}

Here, the Python list comprehensions compute an enumeration of these relations, by iterating over all tuples in the Cartesian product of the argument types and checking a certain condition for each tuple. Alternatively, we can make IDP do this work,  by \verb|Defin|ing the predicates with the appropriate {\em lamba}-expressions.

\begin{lstlisting}
sud.Define("SameRow(Cell, Cell)", 
                 "lambda i, j:  i / 9 == j / 9")
\end{lstlisting}


The cells must be filled with integers from 1 to 9.  It will be convenient to represent an empty cell by the number 0, leading to the following type {\em Number}.
\begin{lstlisting}
sud.Type("Number", range(10))
\end{lstlisting}

We make use of two functions that map cells to numbers: one records the problem statement and the other its solution. The problem statement comes from a list of numbers, that we convert to a dictionary by means of \verb|zip| operation.

\begin{lstlisting}
s=[ 8,5,0, 0,0,2, 4,0,0,   
    7,2,0, 0,0,0, 0,0,9,   
    0,0,4, 0,0,0, 0,0,0,
     ...    ...    ...   ]   
sud.Function("Given(Cell):Number",
              dict(zip(range(len(s)), s)))
sud.Function("Sol(Cell): Number")
\end{lstlisting}

The function \verb|Sol| will be computed by IDP, in accordance with the rules of sudoku.  
First, we state the difference constraint on the appropriate cells.
\begin{lstlisting}
sud.Define("Diff(Cell,Cell)", "lambda x,y: x != y and 
    (SameRow(x,y) or SameCol(x,y) or SameSmallSq(x,y))")
sud.Constraint("all(Sol(x) != Sol(y) for (x,y) in Diff)")
\end{lstlisting}

Next, we state that the solution must match the problem statement on all non-empty cells (i.e., those $\neq 0$), and that it should fill in all cells.
\begin{lstlisting}
sud.Constraint("all(Sol(x) == Given(x) 
                    for x in Cell if Given(x) != 0)")
sud.Constraint("all(Sol(x) != 0 for x in Cell)")
\end{lstlisting}

With this, the sudoku problem is completely specified. The following code passes the \verb|sud.Sol| object to a function that pretty-prints the sudoku.  Only at the start of the \verb|for|-loop in this function is the solution actually computed.

\begin{lstlisting}
def show(grid):
    row = -1
    for x in grid.keys():
        sys.stdout.write(' ' if x % 3 == 0 else '')
        if x % 9 == 0:
            row += 1
            print ('\n' if row % 3 else '')
        print (str(grid[x]) + " "),

show(sud.Sol)
\end{lstlisting}




We remark that, because we use  valid Python expressions to assert constraints, we can use the same expressions to check that the output indeed satisfies the constraints. For instance, at the interactive Python terminal:

\begin{lstlisting}
>>> all(sud.Sol(x) != sud.Sol(y) for (x,y) in sud.Diff)
True
\end{lstlisting}

Peter Norvig has published a sudoku solver written entirely in Python using constraint solving techniques\footnote{\url{http://norvig.com/sudoku.html}}. Not counting whitespace and comments, his code to compute solutions is about 40 lines, whereas the code we have presented in this section is 12 lines.   Moreover, it is easy to replace the existing \verb|solve| function in his code by a call to our knowledge base. This requires two small transformations: first, in Norvig's code, input grids are given in the format of a single string, where an empty cell is represented by a dot; second, he produces output in the form of a dictionary in which the keys are strings of the form $Xn$ with $X \in A..I$ representing the row and $n \in 1..9$ the column.
\begin{lstlisting}
    def solve(sud, grid):
        """ Solve sudoku "grid" using the KB "sud". """
        def translate(n):
            rows     = 'ABCDEFGHI' 
            digits   = '123456789'
            return rows[n/9] + digits[n%9]
        sud.Given = dict(zip(range(81), 
             [int(x) for x in grid.replace('.','0')]))
        return dict(zip(map(translate, sud.Sol.keys()), 
                        map(str, sud.Sol.values())))
\end{lstlisting}

When it comes to runtime, our version is significantly slower than the original on Norvig's test set, averaging $0.27s$ per sudoku versus $0.008s$. 

At the other end of the spectrum, we can also compare to a naive generate-and-test approach. Using Python's powerful \verb|itertools| library, this can also be implemented in about 10 lines of code.  (The code used to test whether a solution is correct can of course use the same syntactical expressions as those which we passed to our API.)  However, the runtime of such a program is very poor: for a sudoku with just 5 empty cells (for comparison, a typical sudoku has around 50 to 60), it already takes over a minute to find a solution. 

Our main conclusions from this experiment are that, at least in this case:
\begin{itemize}
\item Our API can handle, with a limited amount of overhead, the input/output format that a typical Python programmer would use;
\item Our API can be used to develop useful functionality in significantly fewer lines of code (12 versus 40) than a clever Python implementation. In fact, it takes only as many lines of code as a naive generate-and-test algorithm.
\item Even though our API is significantly slower than a clever Python algorithm, it still vastly outperforms the naive generate-and-test approach.
\end{itemize}

\subsection{Working with graphs}

The following class \verb|GraphKB| extends the generic IDP Knowledge Base class with
some specific functionality for working with undirected graphs. When constructing such a \verb|GraphKB|, the nodes of the graph can be initialised by means of a given set and the edges by means of an adjacency list.  The  predicate
\verb|Edge| is \verb|Define|d as the symmetric closure of the adjacency
list.  This class also offers a convenience method to
define the transitive closure of relations over this graph.

\begin{lstlisting}
class GraphKB(IDP):

    def __init__(self, nodes=[0], adj_list=[]):
        super(GraphKB, self).__init__()
        self.Type("Node", nodes)
        self.Predicate("Adjacent(Node,Node)", adj_list)
        self.Define("Edge(Node,Node)", 
            "lambda x,y: Adjacent(x,y) or Adjacent(y,x)")

    def add_TC(self, original, tc_name):
        formula = "lambda x,y: {0}(x,y) or any({0}(x,z) and {0}(z,y) for z in Node)".format(original)
        self.Define(tc_name + "(Node, Node)", formula)
\end{lstlisting}

We can now check if a given adjacency list describes a fully
connected graph:
\begin{lstlisting}
connected = GraphKB(nodes, adj)
connected.define_TC("Path", "Edge")
connected.Constraint("all(Path(x,y) 
                          for x in Node for y in Node)")
if connected.satisfiable:
    print "Graph is fully connected"
\end{lstlisting}

We can use a similar KB to count the number of
connected components in the graph. We do this by selecting a single
representative from each component (its ``\verb|Root|'') and then
counting the number of these representatives.
\begin{lstlisting}
comp = GraphKB(nodes, adj)
comp.define_TC("Path", "Edge")
comp.Predicate("Root(Node)")
comp.Constraint("all(any(Path(r,x) for r in Root) 
                     for x in Node if not Root(x))")
comp.Constraint("not any(Path(x,y) 
                 for x in Root for y in Root if x != y)")
print "Number of components: {0}".format(len(comp.Root))
\end{lstlisting}

For a graph with 1000 nodes in 86 connected components, this program takes 23s to count the components. By comparison, the popular NetworkX Python library\footnote{\url{https://networkx.github.io/}} is two orders of magnitude faster, taking only 0.2s.

In graph theory, an undirected graph is called a {\em tree}
if it is connected and does not contain cycles. When checking for a
cycle in an undirected graph, we of course have to exclude the trivial
two-node cycles that would result from traversing the same undirected
edge in both directions. This in fact makes it easier to use IDP to
check that there {\em is} a cycle, than to check that there {\em is not}
one. The following knowledge base tries to guess the direction in which to
traverse each edge  in order to produce a cycle.  If it is
unsatisfiable, there are no cycles.

\begin{lstlisting}
cyclic = GraphKB()
cyclic.Predicate("Traverse(Node,Node)")
cyclic.Constraint("all(Edge(x,y) for (x,y) in Traverse)")
cyclic.Constraint("not any(Traverse(y,x)
                           for (x,y) in Traverse)")
cyclic.define_TC("TravTC", "Traverse")
cyclic.Constraint("any(TravTC(x,x) for x in Node)")
\end{lstlisting}

We can now combine the two knowledge bases to check whether a given
adjacency list indeed describes a tree.

\begin{lstlisting}
def is_tree(adj_list):
    cyclic.Adjacent = adj_list
    connected.Adjacent = adj_list
    return (bool(connected.satisfiable) 
            and not bool(cyclic.satisfiable))
\end{lstlisting}

This example illustrates how additional functionality can be built on top of the KB objects of our API. In addition, the ability to combine the results of calls to different KBs also allows us to implement functionality that would be harder to implement in a single IDP KB.

\section{Implementation}\label{sec:impl}

The implementation of our API and the examples are available for download.\footnote{\url{https://bitbucket.org/joostv/pyidp/admin}} Interfacing with the IDP system is currently done in a decoupled way: when the API detects that the IDP system needs to be called, it prepares a text file with the appropriate vocabulary, structure and theory, then calles the IDP system as an external process and parses its output. The results of this call are cached, so that the IDP system will not be invoked again until the KB changes.

\section{Related work}\label{sec:rel}

There is already a long history of work attempting to close the gap between imperative and declarative programming \cite{apt98}.  We briefly compare our approach to some recent work in this area.

In \cite{torlak13}, an approach is presented in which a constraint solver is not added to a single host language,  but can be used in the development of a domain-specific language in Racket. Like ours, the motivation behind this work is to allow the power of declarative systems to be more widely used. However, their approach differs, because they count on an intermediary---the designer of the domain-specific language---to hide the complexity of the declarative system, whereas our approach focuses on creating an interface that is natural enough to offer KB functionality directly.

In \cite{koksal12}, a constraint solver is integrated into the Scala language. As ours does, their approach reuses the syntax of the host language to interface with the declarative system. A key difference is that, in their approach, the programmer is explicitly  manipulating, combining and solving constraints, which makes the constraint solver more present in the eventual source code. A second difference is of course that Scala currently appears to be less widely known than Python.

In \cite{Milicevic11}, a reasoner for FO extended with transitive closure is integrated into Java. Their KB language is therefore very similar to (but more restricted than) that of IDP. When it comes to the integration in Java, there are two main differences to our approach. First, the declarative knowledge is not written in expressions in the host language, but in a separate language (the Alloy-like JFSL \cite{yessenov09}). Second, the integration into Java is done in an object-oriented way: the programmer defines classes in which formulas are added as, among others, class invariants, method pre-/postconditions and frame conditions.  In comparison, our Python API seems more lightweight, since it does not require an object-oriented approach. When it comes to computational performance, \cite{Milicevic11} reports good results, which our implementation is not able to match.

In summary, our approach fills the niche of an easy-to-learn quick prototyping API, that, due to Python's current popularity, may speak to a large audience.

\section{Conclusions and future work}

When prototyping an application, a programmer may encounter a computational subproblem for which it would be cumbersome to develop a specific algorithm. The aim of our API is to allow such gaps to be declaratively stopped with as little effort as possible. As we have seen, our API might allow a feasible solution to be produced in only as many lines of code as an (infeasible) naive generate-and-test algorithm. Our use of standard Python objects such as sets and mappings means that no elaborate setup code is required to plug the KB into an existing code base, while our use of standard Python expressions for constraints and definitions leads to a low learning curve.  In addition, both these properties also make it easier to eventually remove the KB if a more efficient solution is required: the same KB that first generated the solution, can later be used to check its correctness, or its constraints may simply be recuperated in the form of Python \verb|assert|-statements.

To prevent changing/removing the KB from leading to code changes elsewhere, our API makes  all calls to the IDP system automatically, whenever they are needed. This has the additional benefit of simplifying the API and not forcing the programmer to learn new terminology. A downside is that it is harder for the programmer to keep track of what is happening when in the program.

Our current implementation of the API is naive in its interfacing with the IDP system, which happens by passing text files (built each time from skratch) to an external process. A better integration, which exploits the Lua interface of IDP, might offer a significant reduction in runtimes. However, since we mainly intend our API to be used in prototyping, this issue might not be pressing. Another consequence, which may be more severe, is that programs written in our API are currently hard to debug: it may be necessary to manually inspect the text file that was passed to the IDP system (in debug-mode, the API  always sends this to standard output). However, this requires the user to be at least somewhat  familiar with IDP input syntax, which is something we aimed to avoid.

Our validation of the API currently consists only of examples that we have implemented ourselves.  A better test would involve Python programmers who have no knowledge of IDP or indeed any declarative system.  However, better debugging facilities seem necessary for such a trial to be successful.

\newpage
\bibliographystyle{plain}
\bibliography{biblio}

\end{document}